\documentclass{article}
\usepackage{spconf,amsmath,epsfig}

\usepackage{soul}
\usepackage{xcolor}
\usepackage{siunitx}
\usepackage{booktabs}
\usepackage{graphicx}
\usepackage{subcaption}
\let\OLDthebibliography\thebibliography
\renewcommand\thebibliography[1]{
  \OLDthebibliography{#1}
  \setlength{\parskip}{0pt}
  \setlength{\itemsep}{0pt plus 0.3ex}
}

\usepackage{enumitem}
\setlist{noitemsep,topsep=0pt, parsep=0pt, partopsep=0pt, leftmargin =*}

\usepackage[numbers,sort&compress,square]{natbib}
\makeatletter
\def\@maketitle{%
  \newpage
  \null
  \vskip 1em%
  \begin{center}%
  \let \footnote \thanks
    {\large \bf \@title \par}%
    \vskip 1.0em%
    {\normalsize
      \lineskip .5em%
      \begin{tabular}[t]{c}%
        \@author
      \end{tabular}\par}%
    \vskip 1em%
  \end{center}%
  \par
  \vskip 1.0em}
\makeatother

\usepackage{authblk}
\author[1]{Youqian~Zhang$^*$\thanks{*These two authors contributed equally.}}
\author[1]{Chunxi~Yang$^*$}
\author[1]{Eugene~Y.~Fu$^\dagger$\thanks{$^\dagger$ Corresponding author. Email:eugene.fu@polyu.edu.hk}\thanks{This work was supported, in part, by PolyU under Grant P0039489, and P0048514.}}
\author[2]{Qinhong~Jiang}
\author[2]{Chen~Yan}
\author[3]{Sze-Yiu~Chau}
\author[1]{Grace~Ngai}
\author[1]{Hong-Va~Leong}
\author[1]{Xiapu~Luo}
\author[2]{Wenyuan~Xu}

\affil[1]{The Hong Kong Polytechnic University, HKSAR} 
\affil[2]{Zhejiang University, Hangzhou, China}
\affil[3]{The Chinese University of Hong Kong, HKSAR}

\begin{document}
\title{Understanding Impacts of Electromagnetic Signal Injection Attacks on Object Detection}

\maketitle


\begin{abstract}
\label{sec:abstract}
Object detection can localize and identify objects in images, and it is extensively employed in critical multimedia applications such as security surveillance and autonomous driving.
Despite the success of existing object detection models, they are often evaluated in ideal scenarios where captured images guarantee the accurate and complete representation of the detecting scenes. However, images captured by image sensors may be affected by different factors in real applications, including cyber-physical attacks.
In particular, attackers can exploit hardware properties within the systems to inject electromagnetic interference so as to manipulate the images. 
Such attacks can cause noisy or incomplete information about the captured scene, leading to incorrect detection results, potentially granting attackers malicious control over critical functions of the systems.
This paper presents a research work that comprehensively quantifies and analyzes the impacts of such attacks on state-of-the-art object detection models in practice. It also sheds light on the underlying reasons for the incorrect detection outcomes.
\end{abstract}

\begin{keywords}
    object detection, image sensor, electromagnetic interference, signal injection attack
\end{keywords}

\section{Introduction}
\label{sec:introduction}

Object detection is a critical computer vision task that involves identifying and locating specific categories of visual objects within images. It has found wide applications in many safety- or security-critical domains, including autonomous driving and surveillance cameras~\cite{zou2023object}. Given the importance of object detection, it is highly likely to be a target for cyber-physical attacks.

In fact, many studies have showcased that achieving attacks on object detection is more than just an imagination. There are various attack modalities that are capable of tampering with image sensors to cause different perturbations onto images. This may lead to incorrect object detection, posing significant threat to safety and security. Taking autonomous driving as an example where a vehicle relies on image sensors to monitor traffic, incorrect detection could result in missing traffic lights or failing to detect another vehicle in front, potentially leading to fatal collisions.

There are two major types of object detection attacks: printing hostile perturbations known as an ``adversarial patch'' onto a target object, and injecting physical signals to manipulate the image sensor.
In recent years, there has been a growing interest in physical signal injection attacks on object detection due to their imperceptibility to the human eye, unlike adversarial patches on objects patches on objects, such as~\cite{duan2020adversarial,zhu2023tpatch,eykholt2018robust}.
The signal injection attack modalities include lights~\cite{petit2015remote, yan2016can, kohler2021they, man2020ghostimage, sayles2021invisible,yan2022rolling, dh2020autonomous}, ultrasounds~\cite{ji2021poltergeist}, and electromagnetic waves~\cite{oyama2021backdoor, kohler2022signal, jiang23glitchhiker}.

\begin{figure}[t]
\centering
\includegraphics[width=0.48\textwidth]{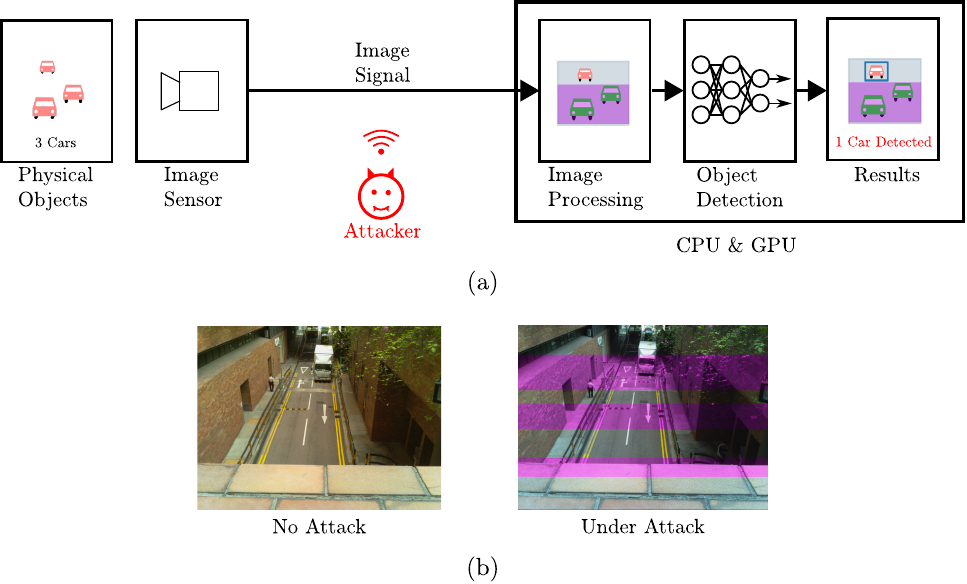}
\caption{(a) An image sensor captures and converts the intensity of light into electrical signals, which are further transmitted to a CPU (and/or GPU) by cables for processing and object detection. Cyber-physical attacks often target on jamming the data transmission to cause perturbations onto images, leading to incorrect object detection. (b) An example under attack image.}
\label{fig:system_model}
\end{figure}

Of particular interest is a recent method called ``GlitchHiker''~\cite{jiang23glitchhiker} that employs electromagnetic waves as the attack modality.
It unveils the principles of how the malicious injection leads to colored strips in images (Fig.~\ref{fig:system_model} b), further affecting the object detection results, for example, preventing detecting objects, as demonstrated in Figure~\ref{fig:system_model}.
What distinguishes this method from earlier approaches is its advantage of not requiring line-of-sight or precise timing knowledge of victim systems.
This characteristic enhances accessibility for potential attackers, thereby indicating a higher potential for abuse in practical scenarios.

However, a comprehensive analysis of the impacts on different object detection models in practical scenes has not been addressed, neither does the underlying reasons for the impacts. 
This work aims to fill these gaps, and the contributions of this work are summarized into three-fold as follows: (1) we construct a practical attack system along with a portable attack device for collecting attacked images in real-world scenarios; (2) we collect, to our best knowledge, the first dataset that include images under attacks in real-world scenarios, in particular street environment; and (3) we conduct comprehensive experiment to quantify and evaluate the impacts of the attacks on different object detection models, unveiling the reasons why some models are robust to the attacks while others are not.




\vspace*{-0.1in}
\section{Experimental Attack Device}
\label{sec:background}

In this work, we particularly investigate attack methodologies that exploit electromagnetic waves to inject malicious signals, i.e., \textit{electromagnetic signal injection attacks (ESIA)}. Based on that, we construct our experimental attack device.

\vspace*{-0.1in}
\subsection{Electromagnetic Signal Injection Attacks}
\label{sec:background_emi_attack}

Cables connecting electronic components in circuits can act like antennas to pick up environmental electromagnetic waves~\cite{wilson2010radiation, paul2022introduction}. Attackers can exploit such ``antenna-like'' property to inject malicious signals into the circuits by radiating electromagnetic waves, and as such, manipulate the waveform of signals.
The injection process can be complex, with many factors impacting its outcome. 
Basically, the effectiveness and efficiency of the injection are determined by the attack power and the attack frequency~\cite{yan2020sok}.
It is demonstrated that the injected signal needs to be strong enough to cause effective impacts; increasing the attack power enhances the likelihood of a successful injection~\cite{zhang2022electromagnetic}.
Regarding frequencies, the attack frequency needs to align with the resonant frequency of the cable to maximize the strength of the injected signals.
A practical way of determining the resonant frequency is to obtain a replica of the target system and then scan frequencies and measure the system's responses~\cite{kune2013ghost}.

After the injection, how the circuits respond to the injected signal is critical for the success of the attacks. In this work, the specific focus is given to the  circuit's response on digital signals, which are represented by sequences of 0s and 1s.
Receiving circuits decide the value of a digital signal by comparing its voltage level with two voltage thresholds
As mentioned previously, the injected signals manipulate the waveform of signals, meaning that they can push the voltage level corresponding to 1 to a new level corresponding to 0, resulting in an incorrect reception of 0, or vice versa.
Such a principle has been demonstrated in many previous studies~\cite{jang2023paralyzing, jiang23glitchhiker, zhang2023electromagnetic, zhang2022electromagnetic, selvaraj2018electromagnetic, dayanikli2022wireless,xie2023bitdance, kohler2023brokenwire}.
To put it simply, \textit{ESIA} can cause bits to be received incorrectly.

\vspace*{-0.1in}
\subsection{Colored Strips Across Images}
\label{sec:background_color_change_in_image}

\begin{figure}[t]
\centering
\includegraphics[width=0.4\textwidth]{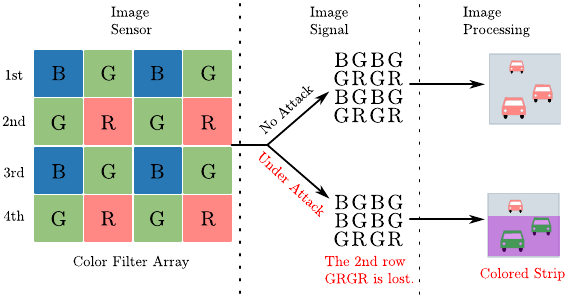}
\caption{Under attack, for example, the second row of pixels is lost, leading to a colored strip in the image.}
\label{fig:colored_strips}
\end{figure}

The image sensor consists of arrays of photodiodes, each of which measures the intensity of the incoming light and converts it into electrical signals, each of which corresponds to a single pixel in the final image. 
To capture color information, every photodiode is overlaid with a color filter, which allows lights of specific wavelengths to pass through. 
The color filters are arranged in a certain pattern~\cite{lukac2005color}, for example, the Bayer filter mosaic~\cite{bayer1976color}, which alternates between red (R), green (G), and blue (B) filters, as shown in Figure~\ref{fig:colored_strips}.
The image signals generated by the image sensor are read out row by row.
After that, image processing is conducted to reconstruct the image. 

\textit{ESIA} can cause incorrect bits in the image signals.
Such bit errors can be spotted by checksum mechanisms, which are commonly applied to verify data integrity in communications.
A previous study~\cite{jiang23glitchhiker} experimentally showed and verified that in many commercial off-the-shelf image sensor products, the bit errors will lead to the discard of the erroneous row, whose positions will be filled by subsequent rows; further, since the image processing process does not handle the missing row properly, colors will be misinterpreted.
As a result, a colored strip appears, as shown in Figure~\ref{fig:colored_strips}.
Note that more non-consecutive rows are lost, and more colored strips will be caused~\cite{jiang23glitchhiker}. This may eventually lead to incorrect detection on the present object detection models.

\begin{figure}[t]
\centering
\includegraphics[width=0.45\textwidth]
{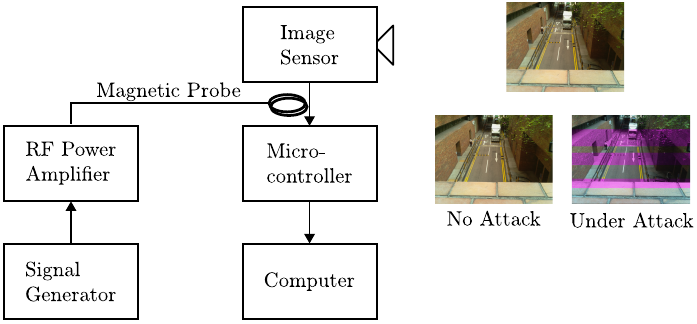}
\caption{An example attack process.}
\label{fig:experiment_setup}
\end{figure}

\vspace*{-0.1in}
\subsection{Attack Device Implementation}

By drawing on the theory and outcome of \textit{ESIA}, we construct a practical attack system along with an attack device to collect images under attacks in real-word scenarios, and understand the impacts on object detection. An \textit{ESIA} process is depicted in Figure~\ref{fig:experiment_setup}.
The object detection system is comprised of an image sensor, a microcontroller, and a computer. The image sensor is connected to the microcontroller via a cable, and the captured images are reconstructed within the microcontroller. 
These reconstructed images are then transmitted to the computer for object detection and analysis.

Our attack device consists of a signal generator, a radio-frequency (RF) power amplifier, and a magnetic probe. 
The signal generator is utilized to generate sinusoidal attacking signals, with a specific attack frequency of \SI{32.5}{\mega\hertz} chosen based on frequency scanning mentioned in Section~\ref{sec:background_emi_attack}. 
These attacking signals are then amplified by the RF power amplifier, which operates at a fixed power level of approximately \SI{3}{\watt}.
The magnetic probe is placed in close proximity (around \SI{2}{\centi\meter}) to the cable that connects the image sensor and the microcontroller, facilitating the injection of the attacking signals at such a low power level.

\vspace*{-0.1in}
\section{Dataset}
\label{sec:dataset}

There is a public available dataset presented in GlitchHiker~\cite{jiang23glitchhiker}, that include under-\textit{ESIA} images. However, its under-attack images are taken by photographing replayed images on a monitor, which are extracted from an open-source traffic photo set~\cite{yu2020bdd100k}. In another word, they are not captured directly from the actual scene. In this work, we collect our own real-world under-attack image sets by employing the attack device and system introduced in Section~\ref{sec:background}, to fully understand impacts of the attacks in real-life applications.

Before running the experiment and data collection, an approval from the host university's IRB is obtained.
The device is placed on a bridge over a road, where access control measures, including rising gates, are implemented to ensure the slow movement of vehicles and pedestrians.
While capturing images, two consecutive photos will be taken each time: one is when no attacking signal is radiated, and the other one is under the influence of the attacking signal.
The interval between taking two photos is minimized as small as possible so as to guarantee two photos are (almost) the same except for the colored strips. Finally, we collected 94 pairs of \textit{Non-Attack} and \textit{Under-Attack} images. 

For the purpose of comparison, we conduct the experiment and analysis on both selected GlitchHiker~\cite{jiang23glitchhiker} dataset (85 pairs of images) and our dataset. Since the GlitchHiker dataset is collected based on replayed images, we denote it as ``REP'' dataset in this paper. On the other hand, we refer to our dataset as ``AUT'' dataset, as it is collected from the authentic scene.
Both datasets were annotated by at least three human experts, with the ground truth annotations established through a majority-vote scheme. We particularly focus on four categories, namely person, car, truck, and bus. 


\vspace{-0.1in}
\section{Object Detection Models}
\label{sec:models}

Object detection is a widely explored task in computer vision, leading to various significantly accurate models, such as those utilizing convolutional neural networks (CNN) and Transformer~\cite{wang2024introduction}.
To ensure a thorough and diverse examination across different object detection model architectures, we particularly select Faster R-CNN~\cite{ren2015faster}, YOLOv8~\cite{yolov8}, and Co-DETR~\cite{zong2023detrs} for the evaluation experiment. These models achieve promising benchmark performances, and are widely used. At the same time, they are built upon three different model architectures respectively:
\begin{itemize}
    \item \textbf{Faster R-CNN~\cite{ren2015faster}} utilizes a two-stage architecture for object detection, involving initial region proposals followed by precise classification and bounding box regression. It is widely applied in many applications requiring high accuracy in localization, such as medical imaging and quality inspection. 

    \item \textbf{YOLOv8~\cite{yolov8}} is the latest and the state-of-the-art iteration in the YOLO series, which employ one-stage architecture for object detection. YOLO models detect the object bounding boxes and classes directly from the images with one forward pass. This contribute to their exceptional speed and real-time detection capabilities. 

    \item \textbf{Co-DETR~\cite{zong2023detrs}} is a transformer-based object detection approach, which achieves the state-of-the-art performance on the COCO benchmark~\cite{hansen2021coco}. 
    Its ability to effectively capture complex dependencies and handle diverse object scales and densities makes it robustness across diverse applications. 

\end{itemize}


In this work, we investigate the impacts of \textit{ESIA} on these models. Specifically, we apply the models pre-trained on a benchmark dataset COCO~\cite{hansen2021coco}, and evaluate how they perform on the under-attack images.

\vspace{-0.1in}
\section{Experiment and Results}

We apply the object detection models (Section~\ref{sec:models}) on the two datasets to quantify the performance disparity between normal and under-attack images. 
This enable the quantitative investigation on the impacts of \textit{ESIA}.

\vspace{-0.1in}
\subsection{Performance Metrics}

We adopt a standard metric in object detection evaluation, namely, mean average precision (mAP) to investigate the performance disparity caused by attacks. Specifically, mAP utilizes intersection over union (IoU) to measure the overlap between the actual and detected object bounding boxes, and jointly consider the object classification precision and recall across different IoU thresholds. It offers a comprehensive measurement of object detection accuracy. Please refer to~\cite{hansen2016coco} for more details.


In addition to standard metric, we also examine the ``specific attack effects'' and ``attack success rate'' in the experiment, to understand the performance degradation caused by \textit{ESIA} more comprehensively. Considering a specific class of objects, incorrect detection can either be a false positive (FP) or false negative (FN). FP refers to incorrect detection of a nonexistent object; in other words, the attack may deceive the object detection algorithm into detecting objects that do not exist, and we call such effects as \textit{creating effects}. FN refers to undetected ground truth, meaning that the attack may prevent the object detection algorithm from detecting objects, and we call such effects as \textit{hiding effects}. 
Furthermore, a successful attack is characterized by the manifestation of one or more predefined effects within a pair of target images. Based on that, we formulate ``attack success rate (SR)'' as $\frac{\text{success\_attack\_images}}{\text{total\_under-attack\_images}}$.

\subsection{Results and Discussion}

Table~\ref{table:map} shows the experimental results.
Notably, there is a consistent decline in object detection performance (i.e., mAP) across all the public pre-trained models, when they are applied on the \textit{Non-Attack} versus \textit{Under-Attack} images.
This suggests that the models, while pre-trained on a robust dataset like COCO, are still vulnerable to \textit{ESIA}, which can significantly compromise their detection accuracy.

\begin{table}[h]
\centering
\small
\vspace{-0.06in}
\caption{mAP evaluation (public pre-trained models).}
\label{table:map}

\begin{tabular}{@{}cccc@{}}
\toprule
\textbf{Dataset} & \textbf{Method} & \textbf{Non-Attack} & \textbf{Under-Attack} \\
\midrule
AUT & Faster & 16.0\% & 10.9\% \\
REP & Faster & 25.4\% & 16.0\% \\
\hline
AUT & YOLOV8 & 26.9\% & 21.0\% \\
REP & YOLOV8 & 28.5\% & 26.8\% \\
\hline
AUT & CoDETR & 31.6\% & 26.8\% \\
REP & CoDETR & 32.4\% & 31.8\% \\
\bottomrule
\end{tabular}

\vspace{-0.05in}
\end{table}

However, it is also noting that the public pre-trained models (pre-trained on COCO dataset) are generally less effective on the two datasets, even for images without attacks. When applying object detection models in real-life applications, it is commonly necessary to adjust the models specifically for the target scenarios, to guarantee their effectiveness in the target scenarios. This is often conducted by fine-tuning the models on the datasets collected from the target scenarios. Therefore, in this work, we also fine-tune the models on the two target datasets (fine-tuning on the \textit{Non-Attack} images) and evaluate the performance degradation caused by the attacks. This enable us to fully understant the impacts of the attacks, especially in a way that is more aligned with real-life applications. Table~\ref{table:mapFT} depicts the results.

Upon fine-tuning, all the models yield satisfactory performance on \textit{Non-Attack} images. Co-DETR particularly shows enhanced resilience. It exhibits the most minor performance degradation under attack conditions. This suggests that Co-DETR, potentially due to its transformer-based design, is more robust to such adversarial influences. YOLOv8 and Faster R-CNN, on the other hand, experience a noticeable decrease in mAP when subjected to attacks, confirming the adversarial effect's potency. Notably, the performance gap post-fine-tuning indicates that while fine-tuning (on the target scenario) can enhance the model efficiency on the target scenarios, it cannot counteract the adversarial perturbations introduced by the electromagnetic signal injection.


\begin{table}[h]
\centering
\small
\vspace{-0.06in}
\caption{mAP evaluation (fine-tuned models).}
\label{table:mapFT}
\begin{tabular}{@{}cccc@{}}
\toprule
\textbf{Dataset} & \textbf{Method} & \textbf{Non-Attack} & \textbf{Under-Attack} \\
\midrule
AUT & Faster & 67.6\% & 50.5\% \\
REP & Faster & 69.9\% & 42.8\% \\
\hline
AUT & YOLOV8 & 68.0\% & 31.1\% \\
REP & YOLOV8 & 81.9\% & 37.3\% \\
\hline
AUT & CoDETR & 90.3\% & 88.4\% \\
REP & CoDETR & 89.1\% & 87.7\% \\
\bottomrule
\end{tabular}
\vspace{-0.05in}
\end{table}

We then look into the detailed FP and FN counts to understand the specific \textit{creating} and \textit{hiding} effects along with the attack success rate (SR), mean precision (MP), and mean recall (MR).
To further refine our analysis, we manually filtered out the FPs and FNs that were not directly attributable to the attack effects. This careful curation ensured that our statistics accurately reflected the impact of the generated attack. This step is crucial for isolating the specific influence of the attack from other potential sources of detection errors, thereby providing a more accurate and relevant understanding of the attack's true impact.

\begin{table*}[ht!]
\centering
\small
\vspace{-0.1in}
\caption{Evaluation of the specific attack effects}
\label{tab:FN_FP_Distribution}
\begin{tabular}{@{}l l cccc cccc c c c@{}}
\toprule
\textbf{Dataset} & \textbf{Model} & \multicolumn{4}{c}{\textbf{FN Count}} & \multicolumn{4}{c}{\textbf{FP Count}} & \textbf{SR} & \textbf{MP} & \textbf{MR} \\
\cmidrule(lr){3-6} \cmidrule(lr){7-10}
 & & \textbf{Person} & \textbf{Car} & \textbf{Truck} & \textbf{Bus} & \textbf{Person} & \textbf{Car} & \textbf{Truck} & \textbf{Bus} & \\
\midrule
AUT & Faster(P) & 35 & 78 & 7 & 3 & 5 & 2 & 10 & 12 & 65.1\% & 66.0\% & 44.2\% \\
AUT & Faster(F) & 21 & 6 & 7 & 1 & 0 & 2 & 0 & 1 & 27.66\% & 91.3\% & 79.3\% \\
\hline
REP & Faster(P) & 3 & 12 & 7 & 1 & 5 & 45 & 10 & 3 & 76.7\% & 65.4\% & 78.3\% \\
REP & Faster(F) & 3 & 26 & 9 & 1 & 0 & 6 & 0 & 0 & 41.18\% & 99.0\% & 73.5\% \\
\hline
AUT & YOLO(P) & 35 & 87 & 24 & 3 & 4 & 2 & 12 & 8 & 85.2\% & 63.0\% & 33.7\% \\
AUT & YOLO(F) & 38 & 39 & 14 & 1 & 0 & 4 & 34 & 0 & 59.57\% & 86.5\% & 63.4\% \\
\hline
REP & YOLO(P) & 10 & 55 & 6 & 3 & 1 & 10 & 11 & 2 & 86.7\% & 54.9\% & 50.4\% \\
REP & YOLO(F) & 6 & 72 & 10 & 6 & 0 & 1 & 0 & 0 & 61.18\% & 99.8\% & 44.1\% \\
\hline
AUT & CoDETR(P) & 17 & 46 & 12 & 1 & 2 & 1 & 5 & 2 & 41.5\% & 83.2\% & 70.7\% \\
AUT & CoDETR(F) & 5 & 1 & 2 & 0 & 0 & 0 & 0 & 1 & 5.32\% & 93.8\% & 96.9\% \\
\hline
REP & CoDETR(P) & 6 & 38 & 4 & 2 & 3 & 8 & 8 & 1 & 34\% & 75.7\% & 68.4\% \\
REP & CoDETR(F) & 0 & 4 & 2 & 0 & 0 & 0 & 0 & 0 & 7.06\% & 100.0\% & 96.6\% \\
\bottomrule
\end{tabular}
\end{table*}

As presented in Table~\ref{tab:FN_FP_Distribution}, a significant reduction in the attack success rate is observed when comparing the public pre-trained (``P'') and fine-tuned models (``F''). Fine-tuning the models leads to increased familiarity with the specific scenarios and objects involved, thereby enhancing the model's robustness against the generated attacks. However, it still cannot completely mitigate the effects of the attacks, underscoring the need for further research into defensive strategies.

\begin{figure}[h!]
\centering
\includegraphics[width=0.15\textwidth]{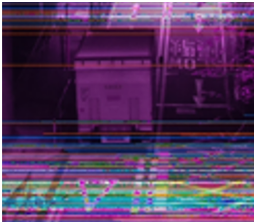}
\caption{CoDETR failed to detect labeled truck}
\label{codetrvis}
\end{figure}

\begin{figure}[h]
    \centering
    \begin{subfigure}[b]{0.325\columnwidth}
        \includegraphics[width=\textwidth]{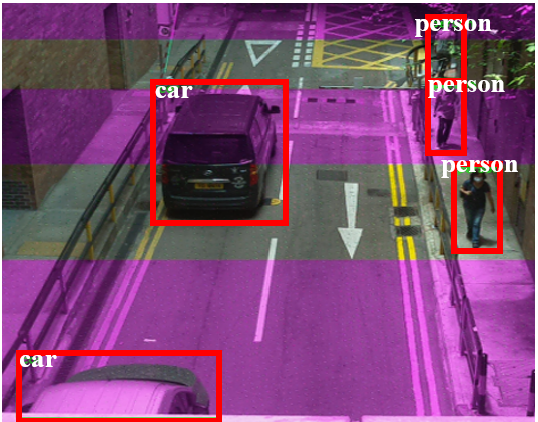}
        \caption{CoDETR}
        \label{fig:sub1}
    \end{subfigure}
    \hfill
    \begin{subfigure}[b]{0.325\columnwidth}
        \includegraphics[width=\textwidth]{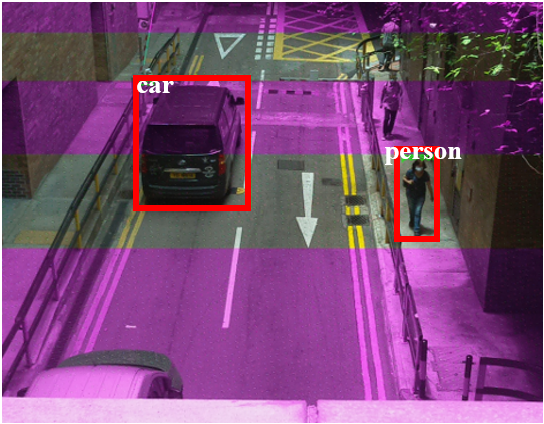}
        \caption{FasterRCNN}
        \label{fig:sub2}
    \end{subfigure}
    \hfill
    \begin{subfigure}[b]{0.325\columnwidth}
        \includegraphics[width=\textwidth]{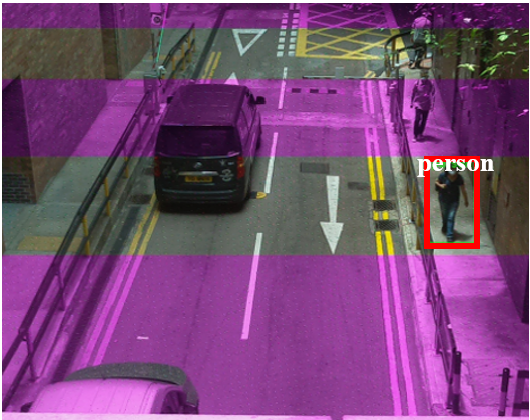}
        \caption{YoloV8}
        \label{fig:sub3}
    \end{subfigure}
    \caption{Detection Result Visualisation}
    \label{fig:test}
\end{figure}

CoDETR consistently exhibits exceptional resilience, maintaining low rates of false negatives (FNs) and false positives (FPs). This performance highlights the robustness of its transformer-based architecture, particularly in response to \textit{ESIA}. Intriguingly, we observed that instances where CoDETR erroneously detected objects in an image were mirrored by similar errors in both FasterRCNN and YoloV8. In many cases, it successfully identifies the annotated objects in images where the other two models failed (Fig~\ref{fig:test}). Although the transformer-based model exhibits a degree of resistance to our attacks, the infrequent but successful breaches, particularly when the attack signals are intense as depicted in Fig ~\ref{codetrvis}, can still substantially undermine the model's detection capabilities, leading to an evident ``hiding effect'' where objects that are successfully detected in normal scene are omitted.



\begin{figure}[h!]
\centering
\includegraphics[width=0.3\textwidth]{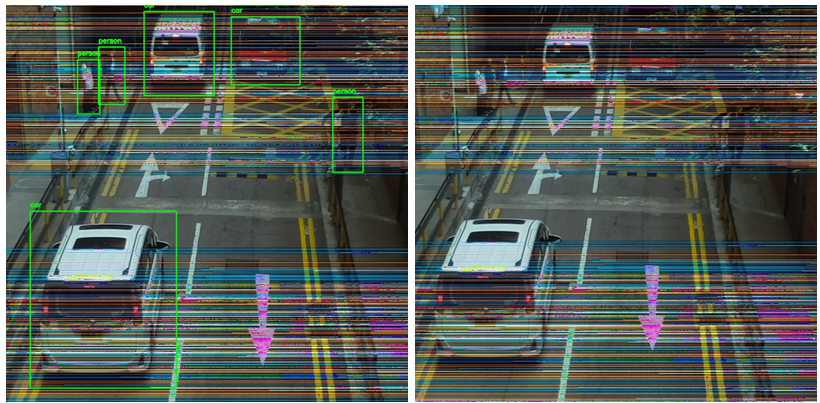}
\caption{FasterRCNN failed to propose ROI for multiple annotated objects }
\label{fastervis}
\end{figure}

In contrast to CoDETR, Faster R-CNN displays a marked vulnerability in detecting smaller objects, with the ``person'' and ``car'' categories being particularly challenging. This issue is most apparent in dataset AUT. The increased difficulty in detecting these objects can be attributed to the camera's elevated position and fixed overhead angle, which tend to reduce the visual size of cars and persons compared to buses and trucks. This phenomenon underscores a pronounced ``hiding'' effect that our attack imposes on Faster R-CNN. The attack patterns interfere with the model's ability to propose regions of interest(ROI), leading to a failure in generating correct predictions for multiple objects (Fig~\ref{fastervis}). This obstruction of the region proposal stage significantly hampers the subsequent detection accuracy, highlighting a critical area of concern for the deployment of two-stage detectors in environments where they might encounter adversarial interference.

\begin{figure}[h!]
\centering
\includegraphics[width=0.2\textwidth]{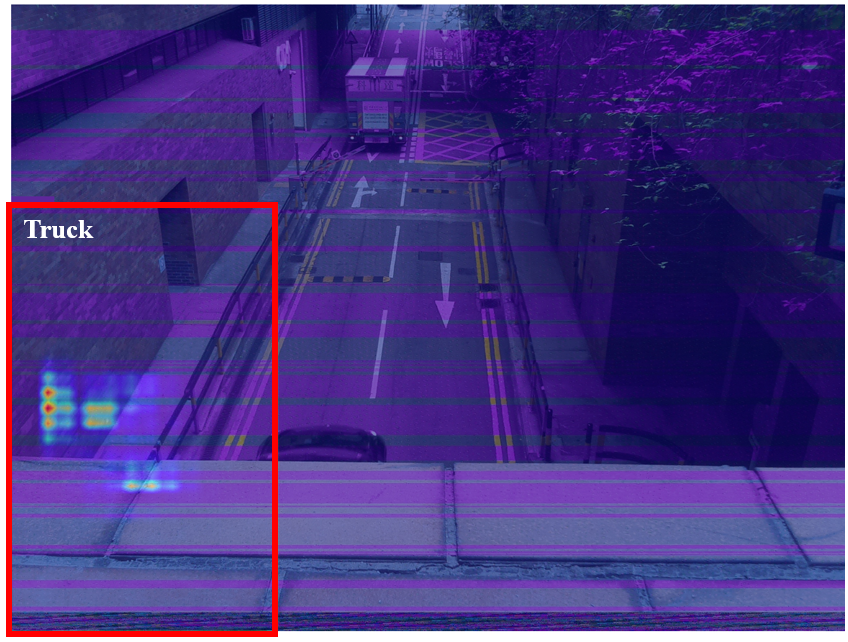}
\caption{YoloV8 GradCAM Vis: False Positive of Class Truck}
\label{fig:yologradcam}
\end{figure}

Meanwhile, YOLOv8's architecture significantly heightens its vulnerability to our attack. Engineered for swiftness and real-time processing, YOLOv8 is a single-shot detector that concurrently processes object localization and classification across the entire image. Although this design is efficient, it may not provide the same level of robustness against the subtle adversarial distortions that the transformer-based CoDETR or the two-stage Faster R-CNN can offer. This inherent fragility, when combined with the nature of overfitting during advanced fine-tuning on standard datasets, renders YOLOv8 particularly prone to misclassifications under adversarial attack conditions. This is evidenced by the excessive false positives for the truck category, where YOLOv8 may mistake large areas of attack-induced visual anomalies for actual features of trucks as shown in~\ref{fig:yologradcam}.


\vspace{-0.1in}
\section{Conclusion}
\label{sec:conclusion}
This study demonstrated the efficacy of electromagnetic signal injection attacks in exploiting vulnerabilities in object detection systems. Our findings reveal that these attacks can significantly disrupt the performance of advanced models, highlighting a critical area of concern in hardware-level security. The research suggests that attackers could potentially refine these techniques to further exploit system weaknesses. This underscores the urgent need for developing robust defense mechanisms to safeguard against such sophisticated attacks.

\bibliographystyle{IEEEbib.bst}
{\footnotesize
\bibliography{mybib}

\begin{thebibliography}{10}

\bibitem{zou2023object}
Zhengxia Zou, Keyan Chen, Zhenwei Shi, Yuhong Guo, and Jieping Ye,
\newblock ``{Object Detection in 20 Years: A Survey},''
\newblock {\em Proceedings of the IEEE}, 2023.

\bibitem{duan2020adversarial}
Ranjie Duan, Xingjun Ma, Yisen Wang, James Bailey, A~Kai Qin, and Yun Yang,
\newblock ``Adversarial camouflage: Hiding physical-world attacks with natural styles,''
\newblock in {\em Proceedings of the IEEE/CVF conference on computer vision and pattern recognition}, 2020, pp. 1000--1008.

\bibitem{zhu2023tpatch}
Wenjun Zhu, Xiaoyu Ji, Yushi Cheng, Shibo Zhang, and Wenyuan Xu,
\newblock ``Tpatch: A triggered physical adversarial patch,''
\newblock in {\em The 32nd USENIX Security Symposium}, 2023.

\bibitem{eykholt2018robust}
Kevin Eykholt, Ivan Evtimov, Earlence Fernandes, Bo~Li, Amir Rahmati, Chaowei Xiao, Atul Prakash, Tadayoshi Kohno, and Dawn Song,
\newblock ``Robust physical-world attacks on deep learning visual classification,''
\newblock in {\em Proceedings of the IEEE conference on computer vision and pattern recognition}, 2018, pp. 1625--1634.

\bibitem{petit2015remote}
Jonathan Petit, Bas Stottelaar, Michael Feiri, and Frank Kargl,
\newblock ``{Remote Attacks on Automated Vehicles Sensors: Experiments on Camera and Lidar},''
\newblock {\em Black Hat Europe}, vol. 11, no. 2015, pp. 995, 2015.

\bibitem{yan2016can}
Chen Yan, Wenyuan Xu, and Jianhao Liu,
\newblock ``{Can You Trust Autonomous Vehicles: Contactless Attacks against Sensors of Self-driving Vehicle},''
\newblock {\em Def Con}, vol. 24, no. 8, pp. 109, 2016.

\bibitem{kohler2021they}
Sebastian K{\"o}hler, Giulio Lovisotto, Simon Birnbach, Richard Baker, and Ivan Martinovic,
\newblock ``They see me rollin’: Inherent vulnerability of the rolling shutter in cmos image sensors,''
\newblock in {\em Annual Computer Security Applications Conference}, 2021, pp. 399--413.

\bibitem{man2020ghostimage}
Yanmao Man, Ming Li, and Ryan Gerdes,
\newblock ``{GhostImage: Remote Perception Attacks against Camera-based Image Classification Systems},''
\newblock in {\em 23rd International Symposium on Research in Attacks, Intrusions and Defenses (RAID 2020)}, 2020, pp. 317--332.

\bibitem{sayles2021invisible}
Athena Sayles, Ashish Hooda, Mohit Gupta, Rahul Chatterjee, and Earlence Fernandes,
\newblock ``{Invisible Perturbations: Physical Adversarial Examples Exploiting the Rolling Shutter Effect},''
\newblock in {\em Proceedings of the IEEE/CVF Conference on Computer Vision and Pattern Recognition}, 2021, pp. 14666--14675.

\bibitem{yan2022rolling}
Chen Yan, Zhijian Xu, Zhanyuan Yin, Stefan Mangard, Xiaoyu Ji, Wenyuan Xu, Kaifa Zhao, Yajin Zhou, Ting Wang, Guofei Gu, et~al.,
\newblock ``Rolling colors: Adversarial laser exploits against traffic light recognition,''
\newblock in {\em 31st USENIX Security Symposium (USENIX Security 22)}, 2022, pp. 1957--1974.

\bibitem{dh2020autonomous}
Sharath~Yadav DH and Asadullah Ansari,
\newblock ``{Autonomous Vehicles Camera Blinding Attack Detection Using Sequence Modelling and Predictive Analytics},''
\newblock Tech. {R}ep., SAE Technical Paper, 2020.

\bibitem{ji2021poltergeist}
Xiaoyu Ji, Yushi Cheng, Yuepeng Zhang, Kai Wang, Chen Yan, Wenyuan Xu, and Kevin Fu,
\newblock ``{Poltergeist: Acoustic Adversarial Machine Learning against Cameras and Computer Vision},''
\newblock in {\em 2021 IEEE Symposium on Security and Privacy (SP)}. IEEE, 2021, pp. 160--175.

\bibitem{oyama2021backdoor}
Tatsuya Oyama, Shunsuke Okura, Kota Yoshida, and Takeshi Fujino,
\newblock ``Backdoor attack on deep neural networks triggered by fault injection attack on image sensor interface,''
\newblock in {\em Proceedings of the 5th Workshop on Attacks and Solutions in Hardware Security}, 2021, pp. 63--72.

\bibitem{kohler2022signal}
Sebastian Kohler, Richard Baker, and Ivan Martinovic,
\newblock ``{Signal Injection Attacks against CCD Image Sensors},''
\newblock in {\em Proc. 2022 ACM ASIA Conference on Computer and Communications Security (ACM ASIACCS 2022)}. 2022, ACM.

\bibitem{jiang23glitchhiker}
Qinhong Jiang, Xiaoyu Ji, Chen Yan, Zhixin Xie, Haina Lou, and Wenyuan Xu,
\newblock ``{GlitchHiker: Uncovering Vulnerabilities of Image Signal Transmission with IEMI},''
\newblock in {\em The 32nd USENIX Security Symposium}, 2023.

\bibitem{wilson2010radiation}
Perry~F Wilson,
\newblock ``{Radiation Patterns of Unintentional Antennas: Estimates, Simulations, and Measurements},''
\newblock in {\em 2010 Asia-Pacific International Symposium on Electromagnetic Compatibility}. IEEE, 2010, pp. 985--989.

\bibitem{paul2022introduction}
Clayton~R Paul, Robert~C Scully, and Mark~A Steffka,
\newblock {\em {Introduction to Electromagnetic Compatibility}},
\newblock John Wiley \& Sons, 2022.

\bibitem{yan2020sok}
Chen Yan, Hocheol Shin, Connor Bolton, Wenyuan Xu, Yongdae Kim, and Kevin Fu,
\newblock ``{Sok: A Minimalist Approach to Formalizing Analog Sensor Security},''
\newblock in {\em 2020 IEEE Symposium on Security and Privacy (SP)}. IEEE, 2020, pp. 233--248.

\bibitem{zhang2022electromagnetic}
Youqian Zhang,
\newblock {\em {Electromagnetic Signal Injection Attacks on Embedded Systems: Modeling and Detection}},
\newblock Ph.D. thesis, University of Oxford, 2022.

\bibitem{kune2013ghost}
Denis~Foo Kune, John Backes, Shane~S Clark, Daniel Kramer, Matthew Reynolds, Kevin Fu, Yongdae Kim, and Wenyuan Xu,
\newblock ``{Ghost Talk: Mitigating EMI Signal Injection Attacks against Analog Sensors},''
\newblock in {\em 2013 IEEE Symposium on Security and Privacy}. IEEE, 2013, pp. 145--159.

\bibitem{jang2023paralyzing}
Joon-Ha Jang, Mangi Cho, Jaehoon Kim, Dongkwan Kim, and Yongdae Kim,
\newblock ``{Paralyzing Drones via EMI Signal Injection on Sensory Communication Channels},''
\newblock in {\em NDSS}, 2023.

\bibitem{zhang2023electromagnetic}
Youqian Zhang and Kasper Rasmussen,
\newblock ``{Electromagnetic Signal Injection Attacks on Differential Signaling},''
\newblock in {\em Proceedings of the 2023 ACM Asia Conference on Computer and Communications Security}, 2023, pp. 314--325.

\bibitem{selvaraj2018electromagnetic}
Jayaprakash Selvaraj, G{\"o}k{\c{c}}en~Y{\i}lmaz Dayan{\i}kl{\i}, Neelam~Prabhu Gaunkar, David Ware, Ryan~M Gerdes, and Mani Mina,
\newblock ``{Electromagnetic Induction Attacks against Embedded Systems},''
\newblock in {\em Proceedings of the 2018 on Asia Conference on Computer and Communications Security}, 2018, pp. 499--510.

\bibitem{dayanikli2022wireless}
G{\"o}k{\c{c}}en~Y{\i}lmaz Dayan{\i}kl{\i}, Abdullah~Zubair Mohammed, Ryan Gerdes, and Mani Mina,
\newblock ``{Wireless Manipulation of Serial Communication},''
\newblock in {\em Proceedings of the 2022 ACM on Asia Conference on Computer and Communications Security}, 2022, pp. 222--236.

\bibitem{xie2023bitdance}
Zhixin Xie, Chen Yan, Xiaoyu Ji, and Wenyuan Xu,
\newblock ``{BitDance: Manipulating UART Serial Communication with IEMI},''
\newblock in {\em Proceedings of the 26th International Symposium on Research in Attacks, Intrusions and Defenses}, 2023, pp. 63--76.

\bibitem{kohler2023brokenwire}
Sebastian K{\"o}hler, Richard Baker, Martin Strohmeier, and Ivan Martinovic,
\newblock ``{Brokenwire: Wireless Disruption of CCS Electric Vehicle Charging},''
\newblock in {\em Proceedings 2023 {{Network}} and {{Distributed System Security Symposium}}}. 2023, {Internet Society}.

\bibitem{lukac2005color}
Rastislav Lukac and Konstantinos~N Plataniotis,
\newblock ``{Color Filter Arrays: Design and Performance Analysis},''
\newblock {\em IEEE Transactions on Consumer Electronics}, vol. 51, no. 4, pp. 1260--1267, 2005.

\bibitem{bayer1976color}
Bryce Bayer,
\newblock ``{Color Imaging Array},''
\newblock {\em United States Patent, no. 3971065}, 1976.

\bibitem{yu2020bdd100k}
Fisher Yu, Haofeng Chen, Xin Wang, Wenqi Xian, Yingying Chen, Fangchen Liu, Vashisht Madhavan, and Trevor Darrell,
\newblock ``Bdd100k: A diverse driving dataset for heterogeneous multitask learning,''
\newblock in {\em Proceedings of the IEEE/CVF conference on computer vision and pattern recognition}, 2020, pp. 2636--2645.

\bibitem{wang2024introduction}
Yuanyuan Wang, Eugene~Yujun Fu, Xinwei Zhai, Chunxi Yang, and Fengchun Pei,
\newblock ``Introduction of artificial intelligence,''
\newblock in {\em Intelligent Building Fire Safety and Smart Firefighting}, pp. 65--97. Springer, 2024.

\bibitem{ren2015faster}
Shaoqing Ren, Kaiming He, Ross Girshick, and Jian Sun,
\newblock ``{Faster R-CNN: Towards Real-time Object Detection with Region Proposal Networks},''
\newblock {\em Advances in neural information processing systems}, vol. 28, 2015.

\bibitem{yolov8}
Glenn Jocher, Ayush Chaurasia, and Jing Qiu,
\newblock ``{YOLO by Ultralytics},'' Jan. 2023.

\bibitem{zong2023detrs}
Zhuofan Zong, Guanglu Song, and Yu~Liu,
\newblock ``{DETRs with Collaborative Hybrid Assignments Training},''
\newblock in {\em Proceedings of the IEEE/CVF international conference on computer vision}, 2023, pp. 6748--6758.

\bibitem{hansen2021coco}
Nikolaus Hansen, Anne Auger, Raymond Ros, Olaf Mersmann, Tea Tu{\v{s}}ar, and Dimo Brockhoff,
\newblock ``Coco: A platform for comparing continuous optimizers in a black-box setting,''
\newblock {\em Optimization Methods and Software}, vol. 36, no. 1, pp. 114--144, 2021.

\bibitem{hansen2016coco}
Nikolaus Hansen, Anne Auger, Dimo Brockhoff, Dejan Tu{\v{s}}ar, and Tea Tu{\v{s}}ar,
\newblock ``Coco: performance assessment,''
\newblock {\em arXiv preprint arXiv:1605.03560}, 2016.

\end{thebibliography}
}

\end{document}